\documentclass[10pt,aps,pra,reprint,showpacs,superscriptaddress]{revtex4-1}
\usepackage[english]{babel}
\usepackage[T1]{fontenc}
\usepackage{natbib}
\usepackage{soul}
\usepackage{amssymb}
\usepackage{bm}
\usepackage{amsmath}
\usepackage[mathcal]{eucal}
\usepackage{color}
\usepackage{graphicx}
\usepackage{hyperref}
\usepackage{mathtools}
\usepackage{url}
\usepackage{tikz}
\usetikzlibrary{shapes}
\usetikzlibrary{decorations}
\usetikzlibrary{arrows}
\usetikzlibrary{matrix}

\newcommand {\nc} {\newcommand}
\nc {\beq} {\begin{eqnarray}}
\nc {\eeqn} [1] {\label{#1} \end{eqnarray}}
\nc {\eoln} [1] {\label{#1} \\}
\nc {\eol} {\nonumber \\}
\nc {\rref} [1] {(\ref{#1})}
\nc {\Eq} [1] {Eq.~(\ref{#1})}
\nc {\Ref} [1] {Ref.~\cite{#1}}
\nc {\la} {\mbox{$\langle$}}
\nc {\ra} {\mbox{$\rangle$}}
\nc {\dem} {\mbox{$\frac{1}{2}$}}
\nc {\cP} {\mathcal{P}}
\nc {\cN} {\mathcal{N}}
\nc {\ve} [1] {\mbox{\boldmath $#1$}}
\nc {\arrow} [2] {\mbox{$\mathop{\rightarrow}\limits_{#1 \rightarrow #2}$}}
\nc {\red}[1] {\textcolor{red}{#1}}
\nc {\mc}[3] {\multicolumn{#1}{#2}{#3}}
\nc {\dd}{\, \mathrm{d}}
\nc {\bs}[1]{\boldsymbol{#1}}
\nc {\ket}[1]{\vert #1 \rangle}
\nc {\bra}[1]{\langle #1 \vert}
\nc {\abs}[1]{\vert #1 \vert}
\nc {\avg}[1]{\langle #1 \rangle}
\nc {\braket}[2]{\langle #1 \vphantom{#2} \vert #2 \vphantom{#1} \rangle}
\nc {\abss}[1]{\left| #1 \right|}

\begin{document}

\title{Multiconfiguration calculations of electronic isotope shift factors in~Al~\textsc{i}}
\author{Livio Filippin}
\email[]{Livio.Filippin@ulb.ac.be}
\affiliation{Chimie Quantique et Photophysique, Universit\'{e} libre de Bruxelles, B-1050 Brussels, Belgium}
\author{Randolf Beerwerth}
\email[]{randolf.beerwerth@uni-jena.de}
\affiliation{Helmholtz-Institut Jena, D-07743 Jena, Germany}
\affiliation{Theoretisch-Physikalisches Institut, Friedrich-Schiller-Universit\"{a}t Jena, D-07743 Jena, Germany}
\author{J\"{o}rgen Ekman}
\email[]{jorgen.ekman@mah.se}
\affiliation{Group for Materials Science and Applied Mathematics, Malm\"{o} University, S-20506 Malm\"{o}, Sweden}
\author{Stephan Fritzsche}
\email[]{s.fritzsche@gsi.de}
\affiliation{Helmholtz-Institut Jena, D-07743 Jena, Germany}
\affiliation{Theoretisch-Physikalisches Institut, Friedrich-Schiller-Universit\"{a}t Jena, D-07743 Jena, Germany}
\author{Michel Godefroid}
\email[]{mrgodef@ulb.ac.be}
\affiliation{Chimie Quantique et Photophysique, Universit\'{e} libre de Bruxelles, B-1050 Brussels, Belgium}
\author{Per J\"{o}nsson}
\email[]{per.jonsson@mah.se}
\affiliation{Group for Materials Science and Applied Mathematics, Malm\"{o} University, S-20506 Malm\"{o}, Sweden}

\date{\today}

\begin{abstract}
The present work reports results from systematic multiconfiguration Dirac-Hartree-Fock calculations of electronic isotope shift factors for a set of transitions between low-lying states in neutral aluminium. These electronic quantities together with observed isotope shifts between different pairs of isotopes provide the changes in mean-square charge radii of the atomic nuclei. Two computational approaches are adopted for the estimation of the mass and field shift factors. Within these approaches, different models for electron correlation are explored in a systematic way to determine a reliable computational strategy and estimate theoretical error bars of the isotope shift factors.
\end{abstract}

\pacs{31.30.Gs, 31.30.jc}

\maketitle

%%%%%%%%%%%%%%%%%%%%%%%%%%%%%%%%%%%%%%%%%%%%%%%%%%%%%%%%%%%%%%%%%%%%%%%%%%%%%%%%%%%%%%%%%%%%%%%%%%%%%%%%%%%%%%%%%%%%%%%%%%
%%%%%%%%%%%%%%%%%%%%%%%%%%%%%%%%%%%%%%%%%%%%%%%%%%%%%%%%%%%%%%%%%%%%%%%%%%%%%%%%%%%%%%%%%%%%%%%%%%%%%%%%%%%%%%%%%%%%%%%%%%

\section{Introduction}
\label{sec:intro}

When the effects of the finite mass and the extended charge distribution of the nucleus are taken into account in a Hamiltonian describing an atomic system, the isotopes of an element display different electronic energy levels~\cite{NGG13}. The isotope shift (IS) of spectral lines, which consists of the mass shift (MS) and the field shift (FS), plays a key role for extracting changes in the mean-square charge radius of the atomic nucleus~\cite{Ki84,CCF12,NLG15}. For a given atomic transition $k$ with frequency~$\nu_k$, it is assumed that the electronic response of the atom to variations in the nuclear mass and charge distribution can be described by only two factors: the mass shift factor, $\Delta K_{k,\text{MS}}$, and the field shift factor, $F_k$, respectively. The observed IS, $\delta \nu_k^{A,A'}$, between any pair of isotopes with mass numbers $A$ and $A'$ is related to the change in nuclear masses and in mean-square charge radii, $\delta \langle r^2 \rangle^{A,A'}$~\cite{Ki84,NGG13}. With this respect, transitions between low-lying levels of neutral aluminium (Al~\textsc{i}) are under investigation in bunched-beam collinear laser spectroscopy experiments~\cite{HMB16} along the Al isotopic chain, in order to determine nuclear properties of the targeted isotopes.

The lack of accurate theoretical calculations of IS in Al~\textsc{i} must be pointed out. Hence, we perform \textit{ab initio} calculations of IS electronic factors using the multiconfiguration Dirac-Hartree-Fock (MCDHF) method, implemented in the \textsc{ris{\small 3}}/\textsc{grasp{\small 2}k}~\cite{NGG13,JGB13} and \textsc{ratip}~\cite{Fr12} program packages. Using the MCDHF method, two different approaches are adopted for the computation of the IS electronic factors in Al~\textsc{i}. The first one is based on the estimation of the expectation values of the one- and two-body recoil Hamiltonian for a given isotope, including relativistic corrections derived by Shabaev~\cite{Sh85,Sh88}, combined with the calculation of the total electron densities at the origin. In the second approach, the relevant factors are extracted from the calculated transition shifts for given triads of isotopes. The results of the two approaches are compared. The same kind of comparison has been performed on neutral copper (Cu~\textsc{i})~\cite{BCF16,CG16}, in order to determine a set of $\delta \langle r^2 \rangle^{65,A'}$ values from the corresponding observed IS.

Very recently, the first computational approach above has been applied to neutral magnesium (Mg~\textsc{i})~\cite{FGE16}, where several transition IS have been determined for the $^{26}$Mg-$^{24}$Mg pair of isotopes. In the present work, the same electron correlation models are applied to Al~\textsc{i}. The second approach was applied to heavier elements such as polonium~\cite{CDS11}, where a good consistency with a King plot analysis was obtained. A similar consistency check was also performed on two transitions in manganese~\cite{HBB16}, where excellent agreement for the mass shift factors computed with \textsc{ris{\small 3}} was observed.

Within both computational approaches, the different correlation models are systematically explored to determine a reliable computational strategy and to estimate theoretical error bars of the IS factors.

In Sec.~\ref{sec:method}, the principles of the MCDHF method are summarised. In Sec.~\ref{sec:isotope_shifts}, the expressions of the MS and FS factors are recalled and the two approaches are further discussed. Section~\ref{sec:AS} enumerates the studied transitions in Al~\textsc{i} and presents the active space expansion strategy adopted for the electron correlation models. In Sec.~\ref{sec:results}, numerical results of the transition energies, as well as of the MS and FS factors, are reported for each studied transition. Section~\ref{sec:conc} reports concluding remarks.

%%%%%%%%%%%%%%%%%%%%%%%%%%%%%%%%%%%%%%%%%%%%%%%%%%%%%%%%%%%%%%%%%%%%%%%%%%%%%%%%%%%%%%%%%%%%%%%%%%%%%%%%%%%%%%%%%%%%%%%%%%
%%%%%%%%%%%%%%%%%%%%%%%%%%%%%%%%%%%%%%%%%%%%%%%%%%%%%%%%%%%%%%%%%%%%%%%%%%%%%%%%%%%%%%%%%%%%%%%%%%%%%%%%%%%%%%%%%%%%%%%%%%

\section{Numerical method}
\label{sec:method}

The MCDHF method~\cite{Gr07}, as implemented in the \textsc{grasp{\small 2}k} program package~\cite{JGB13,JHF07}, is the fully relativistic counterpart of the non-relativistic multiconfiguration Hartree-Fock (MCHF) method~\cite{FTG07,FGB16}. The MCDHF method is employed to obtain wave functions that are referred to as atomic state functions (ASF), i.e., approximate eigenfunctions of the Dirac-Coulomb Hamiltonian given by
\beq
\mathcal{H}_{\text{DC}} = \sum_{i=1}^N [c \, \bs{\alpha}_i \cdot \bs{p}_i + (\beta_i - 1)c^2 + V_{\text{nuc}}(r_i)] + \sum_{i<j}^N \frac{1}{r_{ij}}, \eol
\eeqn{eq_DC_Hamiltonian}
where $V_{\text{nuc}}(r_i)$ is the nuclear potential corresponding to an extended nuclear charge distribution function, $c$ is the speed of light and $\bs{\alpha}$ and $\beta$ are the $(4 \times 4)$ Dirac matrices. An ASF is given as an expansion over $N_{\text{CSFs}}$ $jj$-coupled configuration state functions (CSFs), $\Phi(\gamma_{\nu}\Pi JM_J)$, with the same parity $\Pi$, total angular momentum $J$ and its projection on the $z$-axis, $M_J$:
\beq
\vert \Psi(\gamma\,\Pi JM_J) \rangle = \sum_{\nu=1}^{N_{\text{CSFs}}} c_{\nu} \, \vert \Phi(\gamma_{\nu}\,\Pi JM_J) \rangle.
\eeqn{eq_ASF}

In the MCDHF method, the one-electron radial functions used to construct the CSFs and the expansion coefficients $c_{\nu}$ are determined variationally so as to leave the energy functional
\beq
E = \sum_{\mu,\nu}^{N_{\text{CSFs}}} c_{\mu} c_{\nu} \langle \Phi(\gamma_{\mu}\,\Pi JM_J) \vert \mathcal{H}_{\text{DC}} \vert \Phi(\gamma_{\nu}\,\Pi JM_J) \rangle
\eeqn{eq_energy_functional}
and additional terms for preserving the orthonormality of the radial orbitals stationary with respect to their variations. The resulting coupled radial equations are solved iteratively in the self-consistent field (SCF) procedure. Once radial functions have been determined, a configuration interaction (CI) calculation is performed over the set of configuration states, providing the expansion coefficients for building the potentials of the next iteration. The SCF and CI coupled processes are repeated until convergence of the total wave function~\rref{eq_ASF} and energy~\rref{eq_energy_functional} is reached.

%%%%%%%%%%%%%%%%%%%%%%%%%%%%%%%%%%%%%%%%%%%%%%%%%%%%%%%%%%%%%%%%%%%%%%%%%%%%%%%%%%%%%%%%%%%%%%%%%%%%%%%%%%%%%%%%%%%%%%%%%%
%%%%%%%%%%%%%%%%%%%%%%%%%%%%%%%%%%%%%%%%%%%%%%%%%%%%%%%%%%%%%%%%%%%%%%%%%%%%%%%%%%%%%%%%%%%%%%%%%%%%%%%%%%%%%%%%%%%%%%%%%%

\section{Isotope shift theory}
\label{sec:isotope_shifts}

The finite mass of the nucleus gives rise to a recoil effect that shifts the level energies slightly, called the mass shift (MS). Due to the variation of the IS between the upper and lower levels, the transition IS arises as a difference between the IS for the two levels. Furthermore, the transition frequency MS between two isotopes, $A$ and $A'$, with nuclear masses $M$ and $M'$, is written as the sum of normal mass shift (NMS) and specific mass shift (SMS),
\beq
\delta \nu_{k,\text{MS}}^{A,A'} \equiv \nu_{k,\text{MS}}^{A} - \nu_{k,\text{MS}}^{A'} = \delta \nu_{k,\text{NMS}}^{A,A'} + \delta \nu_{k,\text{SMS}}^{A,A'},
\eeqn{eq_delta_nu_MS}
and can be expressed in terms of a single parameter
\beq
\delta \nu_{k,\text{MS}}^{A,A'} = \left( \frac{1}{M} - \frac{1}{M'} \right) \frac{\Delta K_{k,\text{MS}}}{h} = \left( \frac{1}{M} - \frac{1}{M'} \right) \Delta \tilde{K}_{k,\text{MS}}. \eol
\eeqn{eq_delta_nu_MS_2}
Here, the mass shift factor $\Delta K_{k,\text{MS}}=(K_{\text{MS}}^u-K_{\text{MS}}^l)$ is the difference of the $K_{\text{MS}}=K_{\text{NMS}}+K_{\text{SMS}}$ factors of the upper ($u$) and lower ($l$) levels involved in the transition $k$. For the $\Delta \tilde{K}$ factors, the unit (GHz~u) is often used in the literature. As far as conversion factors are concerned, we use $\Delta K_{k,\text{MS}}\,[m_eE_{\text{h}}]=3609.4824\,\Delta \tilde{K}_{k,\text{MS}}\,[\text{GHz~u}]$.

Neglecting terms of higher order than $\delta \langle r^2 \rangle$ in the Seltzer moment (or nuclear factor)~\cite{Se69}
\beq
\lambda^{A,A'} = \delta \langle r^2 \rangle^{A,A'} + b_1 \delta \langle r^4 \rangle^{A,A'} + b_2 \delta \langle r^6 \rangle^{A,A'} + \cdots, \eol
\eeqn{eq_Seltzer_moment}
the line frequency shift in the transition $k$ arising from the difference in nuclear charge distributions between two isotopes, $A$ and $A'$, can be written as~\cite{FBH95,TFR85,BBP87}
\beq
\delta \nu_{k,\text{FS}}^{A,A'} \equiv \nu_{k,\text{FS}}^{A} - \nu_{k,\text{FS}}^{A'} = F_k \, \delta \langle r^2 \rangle^{A,A'}.
\eeqn{eq_delta_nu_FS}
In the expression above $\delta \langle r^2 \rangle^{A,A'} \equiv \langle r^2 \rangle^{A}-\langle r^2 \rangle^{A'}$ and $F_{k}$ is the electronic factor. Although not used in the current work, it should be mentioned that there are computationally tractable methods to include higher order Seltzer moments in the expression for the transition frequency shift~\cite{EJG16,PCE16}.

The total transition frequency shift is obtained by merely adding the MS, \rref{eq_delta_nu_MS}, and FS, \rref{eq_delta_nu_FS}, contributions:
\beq
\delta \nu_k^{A,A'} & = & \overbrace{\delta \nu_{k,\text{NMS}}^{A,A'} + \delta \nu_{k,\text{SMS}}^{A,A'}}^{\delta \nu_{k,\text{MS}}^{A,A'}} + \, \delta \nu_{k,\text{FS}}^{A,A'} \eol
& = & \left( \frac{1}{M} - \frac{1}{M'} \right) \Delta \tilde{K}_{k,\text{MS}} + F_k \, \delta \langle r^2 \rangle^{A,A'}.
\eeqn{eq_IS}

In this approximation, it is sufficient to describe the total frequency shift between the two isotopes $A$ and $A'$ with only the two electronic parameters given by the mass shift factor $\Delta \tilde{K}_{k,\text{MS}}$ and the field shift factor $F_k$. Furthermore, they relate nuclear properties given by the change in mass and mean-square charge radius to atomic properties. Both factors can be calculated from atomic theory, which is the subject of this work. The two different methods that are applied to compute these quantities are outlined in the next two subsections.

\subsection{Expectation values of the relativistic recoil operator and total electron densities at the origin}
\label{subsec:Ris}

The main ideas of this approach are outlined here and more details can be found in the works by Shabaev \cite{Sh85,Sh88} and Palmer \cite{Pa88}, who pioneered the theory of the relativistic mass shift used in the present work. Gaidamauskas \textit{et al.}~\cite{GNR11} derived the tensorial form of the relativistic recoil operator implemented in \textsc{ris{\small 3}}~\cite{NGG13} and its extension~\cite{EJG16}.

The nuclear recoil corrections within the $(\alpha Z)^4m_e^2/M$ approximation~\cite{Sh85,Sh88} are obtained by evaluating the expectation values of the one- and two-body recoil Hamiltonian for a given isotope,
\beq
\mathcal{H}_{\text{MS}} = \frac{1}{2M} \sum_{i,j}^{N} \left( \bs{p}_i \cdot \bs{p}_j - \frac{\alpha Z}{r_i} \left( \bs{\alpha}_i + \frac{(\bs{\alpha}_i \cdot \bs{r}_i) \bs{r}_i}{r_i^2} \right) \cdot \bs{p}_j \right). \eol
\eeqn{eq_H_MS}
Separating the one-body $(i=j)$ and two-body $(i\neq j)$ terms that, respectively, constitute the NMS and SMS contributions, the Hamiltonian \rref{eq_H_MS} can be written
\beq
\mathcal{H}_{\text{MS}}=\mathcal{H}_{\text{NMS}}+\mathcal{H}_{\text{SMS}}.
\eeqn{eq_H_NMS_SMS} 

The NMS and SMS mass-independent $K$ factors are defined by the following expressions:
\beq
K_{\text{NMS}} \equiv M \langle \Psi \vert \mathcal{H}_{\text{NMS}} \vert \Psi \rangle,
\eeqn{eq_K_NMS}
and
\beq
K_{\text{SMS}} \equiv M \langle \Psi \vert \mathcal{H}_{\text{SMS}} \vert \Psi \rangle.
\eeqn{eq_K_SMS}

Within this approach, the electronic factor $F_{k}$ for the transition $k$ is estimated by
\beq
F_k = \frac{Z}{3\hbar} \left( \frac{e^2}{4\pi \epsilon_0} \right) \Delta \vert \Psi(0) \vert_k^2,
\eeqn{eq_F_k}
which is proportional to the change of the total electron probability density at the origin between the levels $l$ and $u$,
\beq
\Delta \vert \Psi(0) \vert_k^2 = \Delta \rho_k^e(\bs{0}) = \rho_u^e(\bs{0}) - \rho_l^e(\bs{0}).
\eeqn{eq_Delta_Psi_0}

Potential $V_{\text{nuc}}(r_i)$ of \Eq{eq_DC_Hamiltonian} being isotope-dependent, the radial functions vary from one isotope to another, which defines isotopic relaxation. However, the latter is very small and hence neglected along the isotopic chain. Thus, the wave function $\Psi$ is optimized for a specific isotope within this approach.

\subsection{Direct diagonalization of the Hamiltonian matrix}
\label{subsec:Ratip}

Another way to determine $\Delta K_{k,\text{MS}}$ and $F_k$, using an \textit{ab initio} method, is to compute the energies of the upper and lower atomic levels for several isotopes. In this approach we diagonalize the full Hamiltonian matrix including the contribution from the mass shift and the extended nuclear charge distribution~\cite{PFG96,Fr01}, as implemented in \textsc{ratip}~\cite{Fr12}.

For given transition and triad of isotopes $(A,A',A'')$, \Eq{eq_IS} yields a $(2 \times 2)$ system of equations that expresses the computed transition shifts in terms of the unknown IS factors $\Delta K_{k,\text{MS}}$ and $F_k$. Very much resembling the experimental procedure, the system of equations is subsequently solved to obtain the two electronic factors~\cite{Fr12,CCF12}.

This method has the advantage of providing a single set of average mass and field shift factors for a chain of isotopes, adopting the same standard parametrization as for the experimental analysis. Furthermore, the reliability of \Eq{eq_IS} can be estimated by investigating the magnitude of the variations in the calculated factors for different choices of isotope triads. For light and neutral systems like Al~\textsc{i}, this variation is much smaller than the uncertainty due to electron correlation.

The main disadvantage of this method is, that it requires the relativistic CI (RCI) calculations to be performed for a series of selected isotopes. We adopt this approach by first computing the wave functions for the $^{27}$Al isotope, that we use in subsequent RCI calculations for a series of aluminium isotopes to get the transition energies and deduce the corresponding transition IS. For the computations, we used the isotopes $A=19,23,27,31,35$, that cover a wide range of the observed isotopes and the entire range targeted in the planned experiments~\cite{HMB16}. The triads later used for the extraction of the factors are all ten that arise from the above mentioned choice of five isotopes. 

The NMS factor is calculated by including a $\frac{m_e}{M} \sum_i T_i $ term in the Hamiltonian, where $T_i=c \, \bs{\alpha}_i \cdot \bs{p}_i+(\beta_i-1)c^2$ is the Dirac kinetic energy operator associated with electron~$i$~\cite{Pa87,PTF92,LNG12}, which is an approximation of the $\left( \frac{1}{2M} \sum_i p_i^2 \right)$ operator built on the relativistic electron momenta. The SMS operator that is included in the Hamiltonian for the RCI calculations is limited to the standard mass polarization term $ \frac{1}{M} \sum_{i<j} \bs{p}_i \cdot \bs{p}_j $, as described in the write-up of the \textsc{sms{\small 92}} program~\cite{JF97}.

In order to separate the normal and specific mass shift, different calculations have to be carried out. Furthermore, by diagonalization without any mass shift contribution, the field shift factor can be determined independently. This reduces to the direct computation via $F_k = \delta \nu_k/\delta \langle r^2 \rangle$ from \Eq{eq_delta_nu_FS}, if the atomic masses are kept constant. In this particular case, the effect due to the varying nuclear mass along an isotopic chain on the field shift is neglected. However, we did not neglect it in the present calculations, even though a small deviation from the computation via $\delta \nu_k/\delta \langle r^2 \rangle$ was found.

%%%%%%%%%%%%%%%%%%%%%%%%%%%%%%%%%%%%%%%%%%%%%%%%%%%%%%%%%%%%%%%%%%%%%%%%%%%%%%%%%%%%%%%%%%%%%%%%%%%%%%%%%%%%%%%%%%%%%%%%%%
%%%%%%%%%%%%%%%%%%%%%%%%%%%%%%%%%%%%%%%%%%%%%%%%%%%%%%%%%%%%%%%%%%%%%%%%%%%%%%%%%%%%%%%%%%%%%%%%%%%%%%%%%%%%%%%%%%%%%%%%%%

\section{Active space expansion}
\label{sec:AS}

\begin{figure}[ht!]
\begin{center}
\includegraphics[scale=0.9]{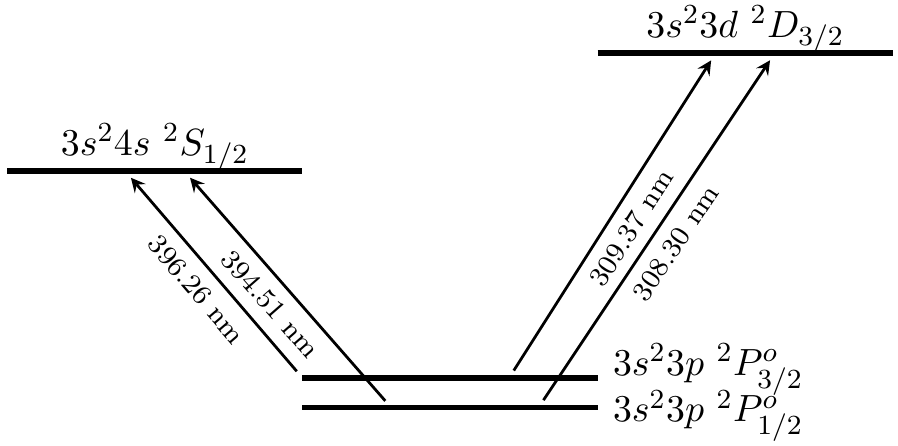}
\end{center}
\caption{\small{Schematic diagram of the Al~\textsc{i} transitions of interest.}}
\label{transitions_diagram}
\end{figure}

Four transitions are under investigation in laser spectroscopy experiments~\cite{HMB16} along the Al isotopic chain, in order to determine nuclear properties of the targeted isotopes (see \figurename{~\ref{transitions_diagram}}): $3s^{2}3p~^{2}P^{o}_{1/2} \rightarrow 3s^{2}4s~^{2}S_{1/2}$ (394.51~nm), $3s^{2}3p~^{2}P^{o}_{3/2} \rightarrow 3s^{2}4s~^{2}S_{1/2}$ (396.26~nm), $3s^{2}3p~^{2}P^{o}_{1/2} \rightarrow 3s^{2}3d~^{2}D_{3/2}$ (308.30~nm) and $3s^{2}3p~^{2}P^{o}_{3/2} \rightarrow 3s^{2}3d~^{2}D_{3/2}$ (309.37~nm).

To effectively capture electron correlation, CSFs of a particular symmetry ($J$) and parity ($\Pi$) are generated through excitations within an active space of orbitals, consisting of orbitals occupied in the reference configurations and correlation orbitals. From hardware and software limitations, it is impossible to use complete active space (CAS) wave functions that would include all CSFs with appropriate $J$ and $\Pi$ for a given orbital active space. Hence the CSF expansions have to be constrained ensuring that major correlation excitations are taken into account~\cite{FGB16}.

Single (S) and double (D) substitutions are performed on a multireference (MR) set, which contains the CSFs that have large expansion coefficients and account for the major correlation effects. These SD-MR substitutions take into account valence-valence (VV), core-valence (CV) as well as core-core (CC) correlations. The VV correlation model only allows SD substitutions from valence orbitals, while the VV+CV correlation model considers SrD substitutions (single and restricted double) from core and valence orbitals, limiting the excitations to a maximum of one hole in the core. By contrast, the VV+CV+CC correlation model allows all SD substitutions from core and valence orbitals.

Within this approach, a \textit{common} orbital basis set is chosen for the lower and upper states of each transition. The reference states are obtained using a valence-CAS procedure: SDT (SD + triple) substitutions are performed within the $n=3,4$ valence orbitals, also including the $5s$ orbital in the active space for the transitions to the $^{2}S_{1/2}$ state (see \tablename{~\ref{table_MR_composition}}). The $5s$ orbital is added to improve the convergence of the $4s$ orbital in the optimisation of the energy functional.

An SCF procedure is then applied to the resulting CSFs, providing the orbital set and the expansion coefficients. Due to limited computer resources, such a valence-CAS MR set would be too large for subsequent calculations when the active orbital space increases. Hence, for reducing the size of the MR set, only the CSFs whose squared expansion coefficients are larger than a given MR cutoff are kept, i.e., $c_\nu^2>\varepsilon_{\text{MR}}$. For each transition, the $\varepsilon_{\text{MR}}$ values and the resulting MR sets are listed in \tablename{~\ref{table_MR_composition}}, for the lower and upper states.

The $1s$ orbital is kept closed in all subsequent calculations, i.e., no substitution from this orbital is allowed. Tests show that opening the $1s$ orbital does not affect the MS and FS factors to any notable extent. Only orbitals occupied in the single configuration DHF approximation are treated as spectroscopic, i.e., are required to have a node structure similar to the corresponding hydrogenic orbitals~\cite{FGB16}. The occupied reference orbitals are frozen in all subsequent calculations. The $J$-levels belonging to a given term are optimised simultaneously with standard weights through the Extended Optimal Level (EOL) scheme~\cite{DGJ89} and the set of virtual orbitals is increased layer by layer.

For a given transition, the optimisation procedure is summarised as follows:

(1) Perform \textit{simultaneous} calculations for the lower state and the upper state of the transition, using an MR set consisting of CSFs with the form $2s^22p^6nln'l'n''l''~J^{\Pi}$ with $n,n',n''=3,4$ (+$5s$) and $l,l',l''=s,p,d,f$. Optimise all orbitals simultaneously. These CSFs account for a fair amount of the VV correlation.

(2) Keep the orbitals fixed from step 1, and optimise an orbital basis layer by layer up to $nl=9h$ for both states of the transition, described by CSFs with respective $J^\Pi$ symmetries. These CSFs are obtained by SD-MR substitutions with the restriction that there is at most one excitation from the $2s^22p^6$ core.

\onecolumngrid

\begin{table}[ht!]
\caption{\small{Reference configurations for the lower and upper states of the studied transitions in Al~\textsc{i}. The MR cutoff values, $\varepsilon_{\text{MR}}$, determine the set of CSFs in the MR space. $N_{\text{CSFs}}$ represents the number of CSFs describing each MR space.}}
\begin{center}
\resizebox{\textwidth}{!}{
\begin{tabular}{l c l l r}
\hline
\hline
\vspace{0.1cm}
Transition                                                  & $\varepsilon_{\text{MR}}$ & $J^\Pi$ & Reference configurations                                                      & $N_{\text{CSFs}}$ \\
\hline
$3s^{2}3p~^{2}P^{o}_{1/2} \rightarrow 3s^{2}4s~^{2}S_{1/2}$ & 0.025                     & $1/2^-$ & [Ne]$\{3s^23p,3s^24p,3s3p3d,3s3d4p,3s4s4p,3p^3,3p^24p,3p3d^2,3s3p5s,3s4p5s\}$ & 14                \\
\vspace{0.1cm}
                                                            &                           & $1/2^+$ & [Ne]$\{3s^24s,3s3p^2,3s3p4p,3s4s^2,3p^24s,3p4s4p,3s4s5s\}$                    & 13                \\
$3s^{2}3p~^{2}P^{o}_{3/2} \rightarrow 3s^{2}4s~^{2}S_{1/2}$ & 0.025                     & $3/2^-$ & [Ne]$\{3s^23p,3s^24p,3s3p3d,3s3d4p,3s4s4p,3p^3,3p^24p,3p3d^2,3s4p5s\}$        & 17                \\
\vspace{0.1cm}
                                                            &                           & $1/2^+$ & [Ne]$\{3s^24s,3s3p^2,3s3p4p,3s4s^2,3p^24s,3p4s4p,3s4s5s,4s4p^2\}$             & 14                \\
$3s^{2}3p~^{2}P^{o}_{1/2} \rightarrow 3s^{2}3d~^{2}D_{3/2}$ & 0.05                      & $1/2^-$ & [Ne]$\{3s^23p,3s3p3d,3s3p4d,3p^3\}$                                           & 7                 \\
\vspace{0.1cm}
                                                            &                           & $3/2^+$ & [Ne]$\{3s^23d,3s^24d,3s3p^2,3s3p4f,3s3d^2,3s3d4d,3s3d4s,3p^23d\}$             & 12                \\
$3s^{2}3p~^{2}P^{o}_{3/2} \rightarrow 3s^{2}3d~^{2}D_{3/2}$ & 0.05                      & $3/2^-$ & [Ne]$\{3s^23p,3s3p3d,3s^24p,3p^3\}$                                           & 7                 \\
\vspace{0.1cm}
                                                            &                           & $3/2^+$ & [Ne]$\{3s^23d,3s^24d,3s3p^2,3s3p4f,3s3d^2,3s3d4d,3s3d4s,3p^23d\}$             & 12                \\
\hline
\hline
\end{tabular}
}
\end{center}
\label{table_MR_composition}
\end{table}

\begin{table}[ht!]
\caption{\small{Level MS factors, $K_{\text{NMS}}$ and $K_{\text{SMS}}$ (in $m_eE_{\text{h}}$), and the electron probability density at the origin, $\rho^e(\bs{0})$ (in $a_0^{-3}$), as functions of the increasing active space for the $3s^{2}3p~^{2}P^{o}_{1/2} \rightarrow 3s^{2}4s~^{2}S_{1/2}$ and $3s^{2}3p~^{2}P^{o}_{3/2} \rightarrow 3s^{2}3d~^{2}D_{3/2}$ transitions in Al~\textsc{i}. Results are computed with \textsc{ris{\scriptsize 3}}. $\Delta^u_l$ stands for the difference between the values of the upper level and the lower level.}}
\begin{center}
\resizebox{\textwidth}{!}{
\begin{tabular}{l c c c c c c c c c c c c c c c c}
\hline
\hline
                &          & & & \mc{3}{c}{$K_{\text{NMS}}$ ($m_eE_{\text{h}}$)} & & & \mc{3}{c}{$K_{\text{SMS}}$ ($m_eE_{\text{h}}$)} & & & \mc{3}{c}{$\rho^e(\bs{0})$ ($a_0^{-3}$)} \\
\cline{5-7} \cline{10-12} \cline{15-17}
\vspace{0.1cm}
Active space    & Notation & & & lower      & upper      & $\Delta^u_l$          & & & lower      & upper      & $\Delta^u_l$          & & & lower       & upper       & $\Delta^u_l$ \\
\hline
\mc{17}{c}{$3s^{2}3p~^{2}P^{o}_{1/2} \rightarrow 3s^{2}4s~^{2}S_{1/2}$}                                                                                                               \\
VV model (MR)   &          & & &            &            &                       & & &            &            &                       & & &             &             &              \\
\vspace{0.1cm}
$5s4p4d4f$      & VV $4f$  & & & $241.9374$ & $241.8043$ & $-0.1331$             & & & $-35.3798$ & $-35.2045$ & $0.1753$              & & & $1497.7637$ & $1498.8441$ & $1.0804$     \\
VV+CV model     &          & & &            &            &                       & & &            &            &                       & & &             &             &              \\
$6s5p5d5f5g$    & CV $5g$  & & & $241.9312$ & $241.8051$ & $-0.1261$             & & & $-35.2342$ & $-35.0512$ & $0.1830$              & & & $1498.0509$ & $1499.2225$ & $1.1716$     \\
$7s6p6d6f6g6h$  & CV $6h$  & & & $241.9528$ & $241.8352$ & $-0.1176$             & & & $-35.2143$ & $-35.0387$ & $0.1756$              & & & $1498.0608$ & $1499.2964$ & $1.2356$     \\
$8s7p7d7f7g7h$  & CV $7h$  & & & $241.9629$ & $241.8378$ & $-0.1251$             & & & $-35.2040$ & $-35.0249$ & $0.1791$              & & & $1498.1162$ & $1499.3015$ & $1.1853$     \\
$9s8p8d8f8g8h$  & CV $8h$  & & & $241.9614$ & $241.8404$ & $-0.1210$             & & & $-35.2032$ & $-35.0225$ & $0.1807$              & & & $1498.1050$ & $1499.3215$ & $1.2165$     \\
\vspace{0.1cm}
$10s9p9d9f9g9h$ & CV $9h$  & & & $241.9629$ & $241.8438$ & $-0.1191$             & & & $-35.2030$ & $-35.0210$ & $0.1820$              & & & $1498.1121$ & $1499.3249$ & $1.2128$     \\
VV+CV+CC model  &          & & &            &            &                       & & &            &            &                       & & &             &             &              \\
\vspace{0.1cm}
$10s9p9d9f9g9h$ & CC $9h$  & & & $242.2185$ & $242.0891$ & $-0.1294$             & & & $-31.5788$ & $-31.4009$ & $0.1779$              & & & $1498.0782$ & $1499.2434$ & $1.1652$     \\
\mc{17}{c}{$3s^{2}3p~^{2}P^{o}_{3/2} \rightarrow 3s^{2}3d~^{2}D_{3/2}$}                                                                                                               \\
VV model (MR)   &          & & &            &            &                       & & &            &            &                       & & &             &             &              \\
\vspace{0.1cm}
$4s4p4d4f$      & VV $4f$  & & & $241.9521$ & $241.7665$ & $-0.1856$             & & & $-35.4048$ & $-35.3722$ & $0.0326$              & & & $1497.6122$ & $1497.5254$ & $-0.0868$    \\
VV+CV model     &          & & &            &            &                       & & &            &            &                       & & &             &             &              \\
$5s5p5d5f5g$    & CV $5g$  & & & $241.9174$ & $241.7866$ & $-0.1308$             & & & $-35.2381$ & $-35.2320$ & $0.0061$              & & & $1497.9986$ & $1497.9804$ & $-0.0182$    \\
$6s6p6d6f6g6h$  & CV $6h$  & & & $241.9459$ & $241.8160$ & $-0.1299$             & & & $-35.2149$ & $-35.2140$ & $0.0009$              & & & $1498.0332$ & $1498.0300$ & $-0.0032$    \\
$7s7p7d7f7g7h$  & CV $7h$  & & & $241.9534$ & $241.8179$ & $-0.1355$             & & & $-35.2053$ & $-35.2033$ & $0.0020$              & & & $1498.0799$ & $1498.0510$ & $-0.0289$    \\
$8s8p8d8f8g8h$  & CV $8h$  & & & $241.9521$ & $241.8151$ & $-0.1370$             & & & $-35.2049$ & $-35.1900$ & $0.0149$              & & & $1498.0782$ & $1498.1535$ & $~~\,0.0753$ \\
\vspace{0.1cm}
$9s9p9d9f9g9h$  & CV $9h$  & & & $241.9539$ & $241.8150$ & $-0.1389$             & & & $-35.2057$ & $-35.1894$ & $0.0163$              & & & $1498.0886$ & $1498.1518$ & $~~\,0.0632$ \\
VV+CV+CC model  &          & & &            &            &                       & & &            &            &                       & & &             &             &              \\
\vspace{0.1cm}
$9s9p9d9f9g9h$  & CC $9h$  & & & $242.1767$ & $242.0466$ & $-0.1301$             & & & $-31.6025$ & $-31.5899$ & $0.0126$              & & & $1497.9558$ & $1497.9980$ & $~~\,0.0422$ \\
\hline
\hline
\end{tabular}
}
\end{center}
\label{table_KNMS_KSMS_rho}
\end{table}

\twocolumngrid

(3) Perform a CI calculation on the CSF expansion with the $J^\Pi$ symmetry of both states, describing VV, CV and CC correlation obtained by SD-MR substitutions to the orbital basis up to $nl=9h$ from step 2.

Following the procedure in steps 1-2 or 1-3 respectively yields results labelled `CV' or `CC' in Tables~\ref{table_E} and \ref{table_KNMS_KSMS_KMS_F}.

The CC effects are more balanced if a common orbital basis is used for describing both the upper and lower states, resulting in more accurate transition energies, as discussed in \Ref{Ve87}.

The CSF expansions become significantly large when CC correlations are taken into account, counting up to $2~\times~10^6$~CSFs. Hence, applying an SCF procedure to such number of CSFs takes too much computing time. This justifies the use of the CI method at that stage.

The effect of adding the Breit interaction to the Dirac-Coulomb Hamiltonian, \rref{eq_DC_Hamiltonian}, is found to be much smaller than the uncertainty in the transition IS factors with respect to the correlation model. This interaction has therefore been neglected in the procedure.

%%%%%%%%%%%%%%%%%%%%%%%%%%%%%%%%%%%%%%%%%%%%%%%%%%%%%%%%%%%%%%%%%%%%%%%%%%%%%%%%%%%%%%%%%%%%%%%%%%%%%%%%%%%%%%%%%%%%%%%%%%
%%%%%%%%%%%%%%%%%%%%%%%%%%%%%%%%%%%%%%%%%%%%%%%%%%%%%%%%%%%%%%%%%%%%%%%%%%%%%%%%%%%%%%%%%%%%%%%%%%%%%%%%%%%%%%%%%%%%%%%%%%

\section{Numerical results}
\label{sec:results}

Let us first study the convergence of the level MS factors, $K_{\text{NMS}}$ and $K_{\text{SMS}}$ (in $m_eE_{\text{h}}$), and the electron probability density at the origin, $\rho^e(\bs{0})$ (in $a_0^{-3}$), of a given transition as a function of the increasing active space. \tablename{~\ref{table_KNMS_KSMS_rho}} displays the values computed with the \textsc{ris{\small 3}} approach for the $3s^{2}3p~^{2}P^{o}_{1/2} \rightarrow 3s^{2}4s~^{2}S_{1/2}$ and $3s^{2}3p~^{2}P^{o}_{3/2} \rightarrow 3s^{2}3d~^{2}D_{3/2}$ transitions.

\vspace{-0.25cm}

\subsection{Valence and core-valence correlations}

For both transitions, the active space is extended within the VV+CV model until convergence of the transition results $\Delta^u_l$ is achieved, which requires the $nl=9h$ correlation layer (`CV~$9h$' in \tablename{~\ref{table_KNMS_KSMS_rho}}). Let us start the analysis with the $3s^{2}3p~^{2}P^{o}_{1/2} \rightarrow 3s^{2}4s~^{2}S_{1/2}$ transition. For $\Delta K_{\text{NMS}}$, adding the orbital layers optimised on VV+CV correlations leads to a change of $10\%$ in comparison with the `VV~$4f$' result. The behaviour is similar for $\Delta K_{\text{SMS}}$ and $\Delta \rho^e(\bs{0})$, where the `CV~$9h$' values differ from the `VV~$4f$' ones by respectively $4\%$ and $12\%$.

The convergence analysis is different concerning the $3s^{2}3p~^{2}P^{o}_{3/2} \rightarrow 3s^{2}3d~^{2}D_{3/2}$ transition. Indeed, from `VV~$4f$' to `CV~$9h$', the $\Delta K_{\text{NMS}}$ value is strongly modified ($25\%$), due to a larger variation of $K_{\text{NMS}}$ for the upper level than for the lower level. This change is even stronger for $\Delta K_{\text{SMS}}$ and $\Delta \rho^e(\bs{0})$, respectively $50\%$ and $173\%$. The fluctuating transition results from `CV~$5g$' to `CV~$7h$' are due to differences of values that are very close to each other for the lower and upper states, and an actual convergence is only achieved at the `CV~$9h$' stage.

A look at the MS and FS factors displayed in \tablename{~\ref{table_KNMS_KSMS_rho}} shows that small variations in the level values due to correlation effects can lead to a significant variation in the transition values, $\Delta^u_l$. This illustrates how sensitive these electronic factors are to the active orbital space used, and hence how challenging it is to obtain reliable values with such a computational approach. This observation also holds for the other transitions studied in this work.

Let us now investigate the agreement of the transition IS factors obtained from the two computational approaches described in Sec.~\ref{sec:isotope_shifts}, i.e., \textsc{ris{\small 3}} and \textsc{ratip}. \tablename{~\ref{table_KNMS_KSMS_KMS_F_CV}} displays the MS factors, $\Delta \tilde{K}_{\text{NMS}}$, $\Delta \tilde{K}_{\text{SMS}}$ and $\Delta \tilde{K}_{\text{MS}}$ (in GHz~u), and the FS factors, $F$ (in MHz/fm$^{2}$), of the studied transitions in Al~\textsc{i} within the VV+CV model.

For each of the two computational approaches, both \textit{common} and \textit{separate} optimisation strategies of the orbital basis sets are considered for the lower and upper states of the transitions. The former strategy corresponds to the one presented in Sec.~\ref{sec:AS}, while the latter strategy implies separate calculations for the lower and upper states, leading to two different orbital basis sets in which orbital relaxation is allowed. Hence, differences in the results of the transition IS factors may arise from two different sources: (i)~discrepancies between \textsc{ris{\small 3}} and \textsc{ratip} approaches considering a given optimisation strategy and (ii)~discrepancies between common and separate optimisation strategies considering a given computational approach. Both sources of discrepancies provide error bars on the IS factors within the VV+CV model.

For both optimisation strategies, a very good consistency is found between the results of \textsc{ris{\small 3}} and \textsc{ratip}, despite the intrinsic differences in the two approaches. Indeed the agreement is not expected to be perfect, since the two approaches do not involve the same operators in the computation of the IS factors. Relativistic corrections to the recoil Hamiltonian~\eqref{eq_H_MS} are part of the explanation for the small discrepancies that are observed.

The major part of the error bars on the IS factors arises from the discrepancies between the two optimisation strategies, whether \textsc{ris{\small 3}} or \textsc{ratip} is used. Concerning the transitions to the $^{2}S_{1/2}$ state, the relative differences stay within $5\%$ for $\Delta \tilde{K}_{\text{MS}}$ and $F$, and are slightly larger for the transitions to the $^{2}D_{3/2}$ state for the reason discussed above.

%\vspace{0.5cm}

\subsection{Core correlations}

Let us go back to \tablename{~\ref{table_KNMS_KSMS_rho}} and analyse the effect of core correlations on the transition IS factors, starting the discussion again with the $3s^{2}3p~^{2}P^{o}_{1/2} \rightarrow 3s^{2}4s~^{2}S_{1/2}$ transition. For $\Delta K_{\text{NMS}}$ and $\Delta K_{\text{SMS}}$, the relative differences from `CV~$9h$' to `CC~$9h$' are respectively $9\%$ and $2\%$, of the same order as the ones within the VV+CV model, while the difference is lower for $\Delta \rho^e(\bs{0})$ ($4\%$ against $12\%$).

Turning to the $3s^{2}3p~^{2}P^{o}_{3/2} \rightarrow 3s^{2}3d~^{2}D_{3/2}$ transition, the relative differences from `CV~$9h$' to `CC~$9h$' reach $6\%$ for $\Delta K_{\text{NMS}}$, $23\%$ for $\Delta K_{\text{SMS}}$ and and $33\%$ for $\Delta \rho^e(\bs{0})$, which is much lower than from `VV~$4f$' to `CV~$9h$'. However, the last two differences are still large, illustrating again the sensitivity of these factors to electron correlation for this transition.

Let us also study the agreement of the transition IS factors obtained with \textsc{ris{\small 3}} and \textsc{ratip} when core correlations are taken into account. \tablename{~\ref{table_KNMS_KSMS_KMS_F_CC}} displays the MS factors, $\Delta \tilde{K}_{\text{NMS}}$, $\Delta \tilde{K}_{\text{SMS}}$ and $\Delta \tilde{K}_{\text{MS}}$ (in GHz~u), and the FS factors, $F$ (in MHz/fm$^{2}$), of the studied transitions in Al~\textsc{i} within the VV+CV+CC model.

For each of the two computational approaches, only a \textit{common} orbital basis set is considered for the lower and upper states of each transition. Treatment with two \textit{separate} orbital bases provides inaccurate transition IS factors in addition to inaccurate transition energies mentioned in Sec.~\ref{sec:AS}, due to the fact that the CC effects~are

\onecolumngrid

%\vspace{0.25cm}

\begin{table}[ht!]
\caption{\small{MS factors, $\Delta \tilde{K}_{\text{NMS}}$, $\Delta \tilde{K}_{\text{SMS}}$ and $\Delta \tilde{K}_{\text{MS}}$ (in GHz~u), and FS factors, $F$ (in MHz/fm$^{2}$), of the studied transitions in Al~\textsc{i} within the VV+CV model. Comparison of the results obtained with \textsc{ris{\scriptsize 3}} and \textsc{ratip}. Both common (`Com.') and separate (`Sep.') orbital basis sets are considered for the lower and upper states of each transition.}}
\begin{center}
\resizebox{\textwidth}{!}{
\begin{tabular}{l c c c c c c c c c c c c c c c c c c c c c c c c c c c c}
\hline
\hline 
                                                            & & & \mc{5}{c}{$\Delta \tilde{K}_{\text{NMS}}$ (GHz~u)} & & & \mc{5}{c}{$\Delta \tilde{K}_{\text{SMS}}$ (GHz~u)} & & & \mc{5}{c}{$\Delta \tilde{K}_{\text{MS}}$ (GHz~u)} & & & \mc{5}{c}{$F$ (MHz/fm$^{2}$)} \\
\cline{4-8} \cline{11-15} \cline{18-22} \cline{25-29}
 & & & \mc{2}{c}{\textsc{ris{\scriptsize 3}}} & & \mc{2}{c}{\textsc{ratip}} & & & \mc{2}{c}{\textsc{ris{\scriptsize 3}}} & & \mc{2}{c}{\textsc{ratip}} & & & \mc{2}{c}{\textsc{ris{\scriptsize 3}}} & & \mc{2}{c}{\textsc{ratip}} & & & \mc{2}{c}{\textsc{ris{\scriptsize 3}}} & & \mc{2}{c}{\textsc{ratip}} \\
\cline{4-5} \cline{7-8} \cline{11-12} \cline{14-15} \cline{18-19} \cline{21-22} \cline{25-26} \cline{28-29}
\vspace{0.1cm}
Transition                                                  & & & Com.   & Sep.    & & Com.   & Sep.   & & & Com.    & Sep.    & & Com.    & Sep.    & & & Com.      & Sep.      & & Com.      & Sep.      & & & Com.     & Sep.     & & Com.     & Sep.     \\
\hline
$3s^{2}3p~^{2}P^{o}_{1/2} \rightarrow 3s^{2}4s~^{2}S_{1/2}$ & & & $-430$ & $-439$  & & $-430$ & $-439$ & & & $657$   & $674$   & & $667$   & $684$   & & & $~~\,227$ & $~~\,235$ & & $~~\,237$ & $~~\,245$ & & & $77.6$   & $74.5$   & & $77.3$   & $78.4$   \\
$3s^{2}3p~^{2}P^{o}_{3/2} \rightarrow 3s^{2}4s~^{2}S_{1/2}$ & & & $-432$ & $-437$  & & $-427$ & $-432$ & & & $656$   & $676$   & & $660$   & $679$   & & & $~~\,224$ & $~~\,239$ & & $~~\,233$ & $~~\,247$ & & & $77.5$   & $74.0$   & & $77.2$   & $78.4$   \\
$3s^{2}3p~^{2}P^{o}_{1/2} \rightarrow 3s^{2}3d~^{2}D_{3/2}$ & & & $-513$ & $-562$  & & $-518$ & $-567$ & & & $~\,58$ & $~\,75$ & & $~\,64$ & $~\,81$ & & & $-455$    & $-487$    & & $-454$    & $-486$    & & & $~\,4.5$ & $~\,4.0$ & & $~\,4.5$ & $~\,4.0$ \\
\vspace{0.1cm}
$3s^{2}3p~^{2}P^{o}_{3/2} \rightarrow 3s^{2}3d~^{2}D_{3/2}$ & & & $-501$ & $-553$  & & $-500$ & $-552$ & & & $~\,59$ & $~\,78$ & & $~\,58$ & $~\,78$ & & & $-442$    & $-475$    & & $-442$    & $-474$    & & & $~\,4.0$ & $~\,3.4$ & & $~\,4.0$ & $~\,3.9$ \\
\hline
\hline
\end{tabular}
}
\end{center}
\label{table_KNMS_KSMS_KMS_F_CV}
\end{table}

\begin{table}[ht!]
\caption{\small{MS factors, $\Delta \tilde{K}_{\text{NMS}}$, $\Delta \tilde{K}_{\text{SMS}}$ and $\Delta \tilde{K}_{\text{MS}}$ (in GHz~u), and FS factors, $F$ (in MHz/fm$^{2}$), of the studied transitions in Al~\textsc{i} within the VV+CV+CC model. Comparison of the results obtained with \textsc{ris{\scriptsize 3}} and \textsc{ratip}. Only a common orbital basis set is considered for the lower and upper states of each transition.}}
\begin{center}
\resizebox{\textwidth}{!}{
\begin{tabular}{l c c c c c c c c c c c c c c c c}
\hline
\hline
           & \hspace{1cm} & \mc{3}{c}{$\Delta \tilde{K}_{\text{NMS}}$ (GHz~u)} & \hspace{0.75cm} & \mc{3}{c}{$\Delta \tilde{K}_{\text{SMS}}$ (GHz~u)} & \hspace{0.75cm} & \mc{3}{c}{$\Delta \tilde{K}_{\text{MS}}$ (GHz~u)} & \hspace{0.75cm} & \mc{3}{c}{$F$ (MHz/fm$^{2}$)} \\
\cline{3-5} \cline{7-9} \cline{11-13} \cline{15-17}
\vspace{0.1cm}
Transition & \hspace{1cm} & \textsc{ris{\scriptsize 3}} & \hspace{0.5cm} & \textsc{ratip} & \hspace{0.75cm} & \textsc{ris{\scriptsize 3}} & \hspace{0.5cm} & \textsc{ratip} & \hspace{0.75cm} & \textsc{ris{\scriptsize 3}} & \hspace{0.5cm} & \textsc{ratip} & \hspace{0.75cm} & \textsc{ris{\scriptsize 3}} & \hspace{0.5cm} & \textsc{ratip} \\
\hline
$3s^{2}3p~^{2}P^{o}_{1/2} \rightarrow 3s^{2}4s~^{2}S_{1/2}$ & \hspace{1cm} & $-467$ & \hspace{0.5cm} & $-467$ & \hspace{0.75cm} & $642$   & \hspace{0.5cm} & $652$   & \hspace{0.75cm} & $~~\,175$ & \hspace{0.5cm} & $~~\,185$ & \hspace{0.75cm} & $74.5$   & \hspace{0.5cm} & $74.3$   \\
$3s^{2}3p~^{2}P^{o}_{3/2} \rightarrow 3s^{2}4s~^{2}S_{1/2}$ & \hspace{1cm} & $-451$ & \hspace{0.5cm} & $-447$ & \hspace{0.75cm} & $649$   & \hspace{0.5cm} & $652$   & \hspace{0.75cm} & $~~\,198$ & \hspace{0.5cm} & $~~\,205$ & \hspace{0.75cm} & $75.2$   & \hspace{0.5cm} & $75.0$   \\
$3s^{2}3p~^{2}P^{o}_{1/2} \rightarrow 3s^{2}3d~^{2}D_{3/2}$ & \hspace{1cm} & $-534$ & \hspace{0.5cm} & $-538$ & \hspace{0.75cm} & $~~~8$  & \hspace{0.5cm} & $~\,14$ & \hspace{0.75cm} & $-526$    & \hspace{0.5cm} & $-524$    & \hspace{0.75cm} & $~\,3.8$ & \hspace{0.5cm} & $~\,3.8$ \\
\vspace{0.1cm}
$3s^{2}3p~^{2}P^{o}_{3/2} \rightarrow 3s^{2}3d~^{2}D_{3/2}$ & \hspace{1cm} & $-470$ & \hspace{0.5cm} & $-469$ & \hspace{0.75cm} & $~\,45$ & \hspace{0.5cm} & $~\,45$ & \hspace{0.75cm} & $-425$    & \hspace{0.5cm} & $-424$    & \hspace{0.75cm} & $~\,2.7$ & \hspace{0.5cm} & $~\,2.7$ \\
\hline
\hline
\end{tabular}
}
\end{center}
\label{table_KNMS_KSMS_KMS_F_CC}
\end{table}

\twocolumngrid

\noindent not balanced between the two sets of orbitals. Hence, differences in the results may only arise from discrepancies between \textsc{ris{\small 3}} and \textsc{ratip}, and provide error bars on the transition IS factors within the VV+CV+CC model.

As for valence and core-valence correlations, a very good consistency is obtained between the results of \textsc{ris{\small 3}} and \textsc{ratip} with, for each of the studied transitions, relative differences staying within $5\%$ for $\Delta \tilde{K}_{\text{MS}}$ and $0.3\%$ for $F$. The error bars within the VV+CV+CC model are systematically smaller than the ones within the VV+CV model, due to the fact that they are only deduced from a comparison between \textsc{ris{\small 3}} and \textsc{ratip}.

\subsection{Comparison and discussion}

Up to now, convergence within a given correlation model has been investigated together with consistency between two computational approaches or between two optimisation strategies. However convergence and consistency obviously do not imply accuracy, simply because the adopted correlation model may not be adequate for the studied properties. Hence, one also needs to compare the obtained results of the transition energies and IS factors with reference values existing in the literature. \tablename{~\ref{table_E}} displays the energies, $\Delta E$ (in cm$^{-1}$), of the studied transitions in Al~\textsc{i}. As mentioned in Sec.~\ref{sec:AS}, the labels `CV' and `CC' respectively correspond to the procedure in steps 1-2 (VV+CV model) or steps 1-3 (VV+CV+CC model). The values of $\Delta E$, obtained with a common optimisation strategy, are compared with theoretical coupled-cluster results from Das \textit{et al.}~\cite{DSP16} and NIST data~\cite{KRR15}. The relative errors with NIST values are $0.5-0.6\%$ at the `CV' stage and $0.01-0.5\%$ at the `CC' stage. The accuracy of the transition energies is thus systematically improved when CC correlations are accounted for. Furthermore, both sets of values are more accurate than the results from \Ref{DSP16}, whose accuracy ranges from $1.6\%$ to $2\%$.

\begin{table}[ht!]
\caption{\small{Energies, $\Delta E$ (in cm$^{-1}$), of the studied transitions in Al~\textsc{i}. Results obtained with a common optimisation strategy. Comparison with other theory~\cite{DSP16} and NIST~\cite{KRR15}.}}
\begin{center}
\resizebox{0.485\textwidth}{!}{
\begin{tabular}{l c l c c c}
\hline
\hline
                                                            & & \mc{4}{c}{$\Delta E$ (cm$^{-1}$)}                           \\
\cline{3-6}
\vspace{0.1cm}
Transition                                                  & & \mc{1}{c}{CV} & CC        & \Ref{DSP16} & NIST~\cite{KRR15} \\
\hline
$3s^{2}3p~^{2}P^{o}_{1/2} \rightarrow 3s^{2}4s~^{2}S_{1/2}$ & & $25\,495$     & $25\,351$ & $24\,943$   & $25\,347.7576$    \\
$3s^{2}3p~^{2}P^{o}_{3/2} \rightarrow 3s^{2}4s~^{2}S_{1/2}$ & & $25\,376$     & $25\,173$ & $24\,730$   & $25\,235.6956$    \\
$3s^{2}3p~^{2}P^{o}_{1/2} \rightarrow 3s^{2}3d~^{2}D_{3/2}$ & & $32\,638$     & $32\,595$ & $33\,038$   & $32\,435.4333$    \\
\vspace{0.1cm}
$3s^{2}3p~^{2}P^{o}_{3/2} \rightarrow 3s^{2}3d~^{2}D_{3/2}$ & & $32\,525$     & $32\,245$ & $32\,826$   & $32\,323.3739$    \\
\hline
\hline
\end{tabular}
}
\end{center}
\label{table_E}
\end{table}

\tablename{~\ref{table_KNMS_KSMS_KMS_F}} displays the MS factors, $\Delta \tilde{K}_{\text{NMS}}$, $\Delta \tilde{K}_{\text{SMS}}$ and $\Delta \tilde{K}_{\text{MS}}$ (in GHz~u), and FS factors, $F$ (in MHz/fm$^{2}$), of the studied transitions in Al~\textsc{i}. The associated error bars are given within each correlation model, with the use of the notation $a(b)$ standing for $a \pm b$, where the values $a$ and $b$ respectively correspond to the half-sum and the half-difference of the two extremal results. These error bars do not have any statistical meaning; they only measure the agreement between different versions of the calculations for a given correlation model.

The NMS factor, $\Delta \tilde{K}_{k,\text{NMS}}$, can be approximated with the scaling law~\cite{HE30,MS82}
\beq
\Delta \tilde{K}_{k,\text{NMS}} \approx -m_e \nu_k^{\text{expt}},
\eeqn{eq_sl1}
where $\nu_k^{\text{expt}}$ is the experimental transition energy of transition $k$, available in the NIST database~\cite{KRR15}. Although only strictly valid in the non-relativistic framework, \Eq{eq_sl1} is used as a reference value since the relativistic effects are expected to be small for $Z=13$. The relativistic corrections to $\Delta \tilde{K}_{\text{NMS}}$ can be deduced with \textsc{ris{\small 3}} by computing the expectation values of the non-relativistic part of the recoil Hamiltonian~\rref{eq_H_MS}. These corrections are of the order of a few percent for the studied transitions in Al~\textsc{i}.

An analysis of the NMS and SMS factors indicates that the `CV' results are more reliable. Indeed, an inspection of the transitions to the $^{2}D_{3/2}$ state shows that core correlations induce changes in $\Delta \tilde{K}_{\text{NMS}}$ and $\Delta \tilde{K}_{\text{SMS}}$ that vary much from one transition to another. The two lower states, $^{2}P^{o}_{1/2}$ and $^{2}P^{o}_{3/2}$, are separated by a small fine-structure splitting. Hence, no strong $J$-dependence of the IS factors is expected to occur between these two

\onecolumngrid

\newpage

\begin{table}[ht!]
\caption{\small{MS factors, $\Delta \tilde{K}_{\text{NMS}}$, $\Delta \tilde{K}_{\text{SMS}}$ and $\Delta \tilde{K}_{\text{MS}}$ (in GHz~u), and FS factors, $F$ (in MHz/fm$^{2}$), of the studied transitions in Al~\textsc{i}, and their associated error bars. Notation $a(b)$ in use stands for $a \pm b$ (see text). $\Delta \tilde{K}_{\text{NMS}}$ is compared with values from the scaling law~\rref{eq_sl1} (`Scal.').}}
\begin{center}
\resizebox{\textwidth}{!}{
\begin{tabular}{l c l l c c c r r c c l r c c r r}
\hline
\hline
                                                            & \hspace{0.5cm} & \mc{3}{c}{$\Delta \tilde{K}_{\text{NMS}}$ (GHz~u)}  & & & \mc{2}{c}{$\Delta \tilde{K}_{\text{SMS}}$ (GHz~u)} & & & \mc{2}{c}{$\Delta \tilde{K}_{\text{MS}}$ (GHz~u)} & & & \mc{2}{c}{$F$ (MHz/fm$^{2}$)} \\
\cline{3-5} \cline{8-9} \cline{12-13} \cline{16-17}
\vspace{0.1cm}
Transition                                                  & \hspace{0.5cm} & \mc{1}{c}{CV} & \mc{1}{c}{CC} & Scal.~\rref{eq_sl1} & & & \mc{1}{c}{CV} & \mc{1}{c}{CC}                      & & & \mc{1}{c}{CV} & \mc{1}{c}{CC}                     & & & \mc{1}{c}{CV} & \mc{1}{c}{CC} \\
\hline
$3s^{2}3p~^{2}P^{o}_{1/2} \rightarrow 3s^{2}4s~^{2}S_{1/2}$ & \hspace{0.5cm} & $-434(5)$     & $-467(0)$     & $-417$              & & & $670(14)$     & $647(5)$                           & & & $~~\,236(9)$  & $180(5)$                          & & & $76.5(2.0)$   & $74.4(0.1)$   \\
$3s^{2}3p~^{2}P^{o}_{3/2} \rightarrow 3s^{2}4s~^{2}S_{1/2}$ & \hspace{0.5cm} & $-432(5)$     & $-449(2)$     & $-415$              & & & $667(12)$     & $650(2)$                           & & & $~~\,235(12)$ & $201(4)$                          & & & $76.2(2.2)$   & $75.1(0.1)$   \\
$3s^{2}3p~^{2}P^{o}_{1/2} \rightarrow 3s^{2}3d~^{2}D_{3/2}$ & \hspace{0.5cm} & $-540(27)$    & $-536(2)$     & $-533$              & & & $69(12)$      & $11(3)$                            & & & $-470(17)$    & $-525(1)$                         & & & $4.2(0.3)$    & $3.8(0.0)$    \\
\vspace{0.1cm}
$3s^{2}3p~^{2}P^{o}_{3/2} \rightarrow 3s^{2}3d~^{2}D_{3/2}$ & \hspace{0.5cm} & $-526(27)$    & $-469(1)$     & $-532$              & & & $68(10)$      & $45(0)$                            & & & $-458(17)$    & $-424(1)$                         & & & $3.7(0.3)$    & $2.7(0.0)$    \\
\hline
\hline
\end{tabular}
}
\end{center}
\label{table_KNMS_KSMS_KMS_F}
\end{table}

\twocolumngrid

\noindent transitions, as shown by results from the scaling law~\rref{eq_sl1} (`Scal.'). This argument is only fulfilled by the `CV' results, and also holds for the transitions to the $^{2}S_{1/2}$ state.

The values of $\Delta \tilde{K}_{\text{NMS}}$ are compared with the scaling law results. The `CV' values are in better agreement with \Eq{eq_sl1} than the `CC' ones for three transitions, while the agreement is comparable for the remaining one, $3s^{2}3p~^{2}P^{o}_{1/2} \rightarrow 3s^{2}3d~^{2}D_{3/2}$. This observation on Al~\textsc{i} contrasts the recent study on Mg~\textsc{i}~\cite{FGE16}, where it has been shown that core correlations improve the accuracy of the NMS factors. A possible explanation of this inconsistency lies in the values of the MR cutoffs considered in this work. Although it has been shown in \Ref{FGE16} that lowering $\varepsilon_{\text{MR}}$ improves the accuracy of the NMS factors, the accuracy of the calculations performed on Al~\textsc{i} is limited by computer resources. Hence, considering larger MR sets would be too time consuming.

Unlike the NMS factor, no comparison of the computed SMS and FS factors is possible with reference values from state-of-the-art atomic calculations. To our knowledge, no \textit{ab initio} study of IS electronic factors in Al~\textsc{i} is available. From the experimental point of view, Refs.~\cite{CBC96} and \cite{LBC97} report measurements of total IS between $^{26}$Al and $^{27}$Al for the transitions to the $~^{2}D_{3/2}$ state (adopting the sign conventions \rref{eq_delta_nu_MS} and \rref{eq_delta_nu_FS} of the present work): $\delta \nu^{26,27} \equiv \nu^{26} - \nu^{27} = -616(3)$\,MHz for $3s^{2}3p~^{2}P^{o}_{1/2} \rightarrow 3s^{2}3d~^{2}D_{3/2}$ and $-613(1)$\,MHz for $3s^{2}3p~^{2}P^{o}_{3/2} \rightarrow 3s^{2}3d~^{2}D_{3/2}$, where the numbers in parentheses correspond to systematic uncertainties. The FS contribution is estimated to be very small, less than $-7$\,MHz~\cite{CBC96}. Multiplying the `CV' results of $\Delta \tilde{K}_{\text{MS}}$ by $(1/M_{26}-1/M_{27})$ yields $-669(14)$\,MHz and $-652(25)$\,MHz, adopting the same meaning for the error bars as in \tablename{~\ref{table_KNMS_KSMS_KMS_F}}. Comparison between theory and experiment shows that they do not agree within the error bars, although the discrepancies are not large.

Subtracting from the total MS the NMS contribution given by the scaling law
\beq
\delta \nu_{k,\text{NMS}}^{26,27} \approx \left( \frac{m_e}{M_{27}} - \frac{m_e}{M_{26}} \right) \nu_k^{\text{expt}},
\eeqn{eq_sl2}
yields the SMS contribution: $\delta \nu^{26,27}_{\text{SMS}} = 141(8)$\,MHz and $\delta \nu^{26,27}_{\text{SMS}} = 140(8)$\,MHz. Multiplying the `CV' results of $\Delta \tilde{K}_{\text{SMS}}$ by $(1/M_{26}-1/M_{27})$ yields $98(17)$\,MHz and $97(14)$\,MHz, and the comparison shows the same conclusion.

Turning to the FS factors, it is seen that core correlations do not significantly affect the $F$ values for the transitions to the $~^{2}S_{1/2}$ state, where the `CC' results lie within the error bars of the `CV' ones. This is not the case for the two other transitions. Nevertheless, these two transitions are not relevant for future experiments, due to their very low $F$ values. As $1/F$ is used by experimentalists in a King plot technique as the slope of a linear fit, a small error in these $F$ values can induce a large error in the slope, leading to inaccurate results. With this respect, a good subject for laser spectroscopy experiments along the Al isotopic chain would be the study of the $3s^{2}3p~^{2}P^{o}_{J} \rightarrow 3s3p^{2}~^{4}P_{J}$ intercombination transitions in Al~\textsc{i}, since the FS factor is much larger for these transitions than for the other ones, due to the different occupations of the $3s$ orbital.

%%%%%%%%%%%%%%%%%%%%%%%%%%%%%%%%%%%%%%%%%%%%%%%%%%%%%%%%%%%%%%%%%%%%%%%%%%%%%%%%%%%%%%%%%%%%%%%%%%%%%%%%%%%%%%%%%%%%%%%%%%
%%%%%%%%%%%%%%%%%%%%%%%%%%%%%%%%%%%%%%%%%%%%%%%%%%%%%%%%%%%%%%%%%%%%%%%%%%%%%%%%%%%%%%%%%%%%%%%%%%%%%%%%%%%%%%%%%%%%%%%%%%

\section{Conclusion}
\label{sec:conc}

This work describes an \textit{ab initio} method for the relativistic calculation of the IS electronic factors in many-electron atoms using the MCDHF approach. Two computational approaches are adopted for the estimation of the MS and FS factors for transitions between low-lying states in Al~\textsc{i}. The first one, implemented in \textsc{ris{\small 3}}, is based on the expectation values of the relativistic recoil Hamiltonian for a given isotope, together with the field shift factors estimated from the total electron densities at the origin. The second one, implemented in \textsc{ratip}, consists in extracting the relevant factors from the calculated transition shifts for given triads of isotopes. In both of them, different correlation models are explored in a systematic way to determine a reliable computational strategy and estimate theoretical error bars. Results obtained with \textsc{ris{\small 3}} and \textsc{ratip} agree well with each other, since the relativistic corrections to the recoil operator, implemented differently in these two codes, are expected to be small for Al~\textsc{i}.

Within each correlation model, the convergence of the level MS factors and the electronic probability density at the origin, as a function of the increasing active space, is studied for the $3s^{2}3p~^{2}P^{o}_{1/2} \rightarrow 3s^{2}4s~^{2}S_{1/2}$ and $3s^{2}3p~^{2}P^{o}_{3/2} \rightarrow 3s^{2}3d~^{2}D_{3/2}$ transitions. It is shown that small variations in the level values due to correlation effects can lead to a significant variation in the transition values, more pronounced in the latter transition. This observation highlights the challenge in obtaining accurate IS factors with such an approach.

The study performed on Al~\textsc{i} shows that CC correlations need to be accounted for in the computational strategy in order to obtain more accurate values for the transition energies. By contrast, the accuracy of the NMS factors in comparison with results from the scaling law is not improved when CC effects are added, which is in contrast to a similar work performed on Mg~\textsc{i}~\cite{FGE16}. Decreasing the MR cutoff further is possible in Mg~\textsc{i} due to a smaller restricted CSF space, but impossible in Al~\textsc{i} due to computer limitations. Furthermore, both works show that the SMS factors are less accurate within the VV+CV+CC model, while no significant change in the FS factors is found. Hence, the most reliable correlation model remains the VV+CV model.

A possible way to improve the accuracy of the present results is the use of the partitioned correlation function interaction (PCFI) approach~\cite{VRJ13}. It is based on the idea of relaxing the orthonormality restriction on the orbital basis, and breaking down the very large calculations in the traditional multiconfiguration methods into a series of smaller parallel calculations. This method is very flexible for targeting different electron correlation effects. CC effects in IS factors could be then treated more accurately and efficiently with the use of this technique. Additionally, electron correlation effects beyond the SD-MR model (such as triple and quadruple excitations) can be included perturbatively. Work is being done in these directions.

%%%%%%%%%%%%%%%%%%%%%%%%%%%%%%%%%%%%%%%%%%%%%%%%%%%%%%%%%%%%%%%%%%%%%%%%%%%%%%%%%%%%%%%%%%%%%%%%%%%%%%%%%%%%%%%%%%%%%%%%%%
%%%%%%%%%%%%%%%%%%%%%%%%%%%%%%%%%%%%%%%%%%%%%%%%%%%%%%%%%%%%%%%%%%%%%%%%%%%%%%%%%%%%%%%%%%%%%%%%%%%%%%%%%%%%%%%%%%%%%%%%%%

\vspace{-0.5cm}

\begin{acknowledgments}
This work has been partially supported by the Belgian F.R.S.-FNRS Fonds de la Recherche Scientifique (CDR J.0047.16), the BriX IAP Research Program No. P7/12 (Belgium) and by the German Ministry for Education and Research (BMBF) under contract 05P15SJCIA. L.F. acknowledges the support from the FRIA. J.E. and P.J. acknowledge financial support from the Swedish Research Council (VR), under contract 2015-04842.
\end{acknowledgments}

%%%%%%%%%%%%%%%%%%%%%%%%%%%%%%%%%%%%%%%%%%%%%%%%%%%%%%%%%%%%%%%%%%%%%%%%%%%%%%%%%%%%%%%%%%%%%%%%%%%%%%%%%%%%%%%%%%%%%%%%%%
%%%%%%%%%%%%%%%%%%%%%%%%%%%%%%%%%%%%%%%%%%%%%%%%%%%%%%%%%%%%%%%%%%%%%%%%%%%%%%%%%%%%%%%%%%%%%%%%%%%%%%%%%%%%%%%%%%%%%%%%%%

\end{document}